# Analysis of weak elasticity of nematic solids using a continuum approach

Arkady I. Leonov [a1]

*Department of Polymer Engineering, The University of Akron, Akron, OH 44325-0301, USA*

Valery S. Volkov [b]

*Institute of Petrochemical Synthesis, Russian Academy of Sciences., 29 Leninsky Prospect, Moscow119991 , Russia*

Absract. – This paper formally analyses effects of nematic weak elasticity using the five parametric de Gennes' (DG) potential. The analysis is trivialized in a specific (local) Cartesian coordinate system whose one axis is directed along the initial director. We demonstrate that depending on closeness of material parameters to the marginal stability conditions, DG potential describes the entire variety of soft, semi-soft and harder behaviours of weakly elastic nematic solids. We found that along with known shearing soft modes, additional soft stretching mode acting along the director may also exist. For stress tensor components, we proved a theorem of special rotational invariance that is closed to rotational invariance principle postulated by Olmsted. When the only shearing modes are soft, stress tensor is symmetric and DG potential scaled with isotropic shear modulus is reduced to a one-parametric, only strain dependent nematic potential. When additionally the stretching mode is also soft, this dimensionless, strain dependent nematic potential has no additional parameters and shows that the free energy is always less than that in isotropic phase.

## 1. Introduction

Several continuum theories involved internal rotations in elastic free energy to analyse macroscopic elasticity of LC elastomers. Using simple symmetry arguments, de Gennes [1] first introduced the internal rotations in the free energy density for nematic elastomers (see also [2]). Many other researchers [3-9] studied various effects in nematic elastomers at continuum level. Several molecular theories (e.g. see [10-16]) have also been elaborated.

Warner and co-workers [16] first theoretically described the soft modes in nematic elastomers that cost no elastic energy when accompanied by director rotation (e.g. see [15,17]). Earlier, Golubovich and Lubensky [18] generally predicted the existence of soft modes for a broad class of nematic materials as a consequence of their marginal stability. Theoretical predictions of these striking effects with using either dG or Warner potentials were later confirmed experimentally (e.g. see [15]).

Recent review paper [19] attempted to describe all the previous results in terms of symmetries and standard approaches of continuum mechanics.

In spite of the above success of theory for nematic elastomers there are other nematic polymers that demonstrate a semi-soft or even more hard behaviours, which cannot be described by the Warner potential, even in the linear limit. Thus a complete analysis of linear situation within a frame of a more general DG potential [1] seems to be of interest. This is the objective of the present paper.

## 2. Kinematics, free energy and stresses.

For elastic solids of nematic type, an algebraic kinematical relation between the initial value of director $\underline{n}$ in non-deformed state and its actual value $\underline{n}_d$ in the deformed state has the form of orthogonal transformation, with an orthogonal tensor $\underline{\underline{R}}$ :

$$\underline{n}_d = \underline{\underline{R}} \cdot \underline{n} \; ; \qquad \underline{\underline{R}} = \exp(-\underline{\underline{\omega}}^I) \; . \qquad (2.1)$$

Here $\underline{\underline{\omega}}^I$ is the anti-symmetric tensor of finite internal rotations, related to the vector of internal rotations $\underline{\omega}^I$ as: $\omega_{ik}^I = -\delta_{ijk}\omega_j^I$ . In the linear case when the internal rotations are small enough, relations (2.1) yield:

$$\underline{n}_d - \underline{n} \approx -\underline{\underline{\omega}}^I \cdot \underline{n} = \underline{\omega}^I \times \underline{n} \; ; \quad \left|\underline{\underline{\omega}}^I\right| = \left|\underline{\omega}^I\right| <<1 \; . \qquad (2.2)$$

For nematic elastic solids, equations (2.1), (2.2) are complemented by the well-known equations of dynamics of internal rotations (e.g. see [20,21]) and formulation of free energy function.

The state variables in this theory are represented by the tensors of infinitesimal strain $\underline{\underline{e}}$ and relative rotations $\underline{\underline{\omega}} = \underline{\underline{\omega}}^B - \underline{\underline{\omega}}^I$ . Here $\underline{\underline{\omega}}^B$ is the tensor of body rotations, and tensors $\underline{\underline{e}}$ and $\underline{\underline{\omega}}^B$ are represented via the displacement vector $\underline{u}$ by the common formulae of linear elasticity, $2\underline{\underline{e}} = \underline{\nabla}\underline{u} + (\underline{\nabla}\underline{u})^T$; $2\underline{\underline{\omega}}^B = \underline{\nabla}\underline{u} - (\underline{\nabla}\underline{u})^T$. To simplify the analysis the incompressible case, $tr\underline{\underline{e}} = 0$, is considered. The nematic anisotropy is described by the given value $\underline{n}$ of director in undeformed state.

Using simple invariance arguments de Gennes [1] introduced a potential that describes the weak nematic elasticity by the deformation part of free energy $f$ as:

$$2f = \mu_0(\underline{n}\underline{n}:\underline{\underline{e}})^2 + \mu_1(\underline{n}\times\underline{\underline{e}}\times\underline{n})^2 \; + \mu_2(\underline{n}\cdot\underline{\underline{e}}\times\underline{n})^2 \; +\alpha_1(\underline{\underline{\omega}}\times\underline{n})^2 \; +\alpha_2\underline{n}\cdot\underline{\underline{e}}\cdot\underline{\underline{\omega}}\times\underline{n} \; . \qquad (2.3)$$

Here $\mu_k, \alpha_k$ are five material parameters having sense of "moduli". The formulae,

$$(\underline{n} \times \underline{\underline{e}} \times \underline{n})_s = \delta_{ejk} n_j e_{ke} n_m \delta_{mes} ; \quad (\underline{n} \cdot \underline{\underline{e}} \times \underline{n})_s = n_k e_{kt} n_m \delta_{mts} ,$$

reduce (2.3) to the equivalent form of sum of basis tensor invariants, more convenient for our analysis:

$$f = 1/2 G_0 \left| \underline{\underline{e}} \right|^2 + G_1 \underline{nn} : \underline{\underline{e}}^2 + G_2 (\underline{nn} : \underline{\underline{e}})^2 - 2G_3 \underline{nn} : (\underline{\underline{e}} \cdot \underline{\underline{\omega}}) - G_4 \underline{nn} : \underline{\underline{\omega}}^2 . \quad (2.4)$$

Here

$$G_0 = \mu_1, \; G_1 = \mu_2/2 - \mu_1, \; G_2 = (\mu_0 + \mu_1 - \mu_2)/2, \; G_3 = \alpha_2/4, \; G_4 = \alpha_1/2 . \quad (2.5)$$

The material constants in (2.4) depend on the scalar order parameter $Q_0$ in non-strained state. Thus when passing (if possible) through the order-disorder transition to isotropic state, the "nematic" moduli $G_k$ ( $k = 1, 2, 3, 4$ ) disappear but isotropic one $G_0$ does not. In the strained state, the value of scalar order parameter $Q$ can be different from $Q_0$. The change in the scalar order parameter due to deformations was found [22] as $Q - Q_0 \sim (\underline{nn} : \underline{\underline{e}})^2$ and therefore we consider this effect as already described by the parameter $G_2$ in (2.5).

The constitutive equations (CE's) for symmetric, $\underline{\underline{\sigma}}^s$, and anti-symmetric, $\underline{\underline{\sigma}}^a$, parts of extra stress tensor calculated using the form (2.4) of DG potential are:

$$\underline{\underline{\sigma}}^s = \partial f / \partial \underline{\underline{e}} = G_0 \underline{\underline{e}} + G_1 (\underline{nn} \cdot \underline{\underline{e}} + \underline{\underline{e}} \cdot \underline{nn}) + 2G_2 \underline{nn} (\underline{\underline{e}} : \underline{nn}) + G_3 (\underline{nn} \cdot \underline{\underline{\omega}} - \underline{\underline{\omega}} \cdot \underline{nn}) \quad (2.6a)$$

$$\underline{\underline{\sigma}}^a = \partial f / \partial \underline{\underline{\omega}} = G_3 (\underline{nn} \cdot \underline{\underline{e}} - \underline{\underline{e}} \cdot \underline{nn}) + G_4 (\underline{nn} \cdot \underline{\underline{\omega}} + \underline{\underline{\omega}} \cdot \underline{nn}) . \quad (2.6b)$$

The full stress tensor is: $\underline{\underline{\sigma}} = -p\underline{\underline{\delta}} + \underline{\underline{\sigma}}^s + \underline{\underline{\sigma}}^a$, where $p$ is an isotropic pressure.

The material parameters $G_k$ in relation (2.4) present the scaling factors for independent basis tensor invariants. Therefore the following conditions,

$$G_k \neq 0 \; (k = 0, 1, 2, 3, 4) , \quad (2.7)$$

are used to avoid degeneration of equations (2.4), (2.6).

To take into account the smallness of deformations and internal rotations when comparing order values of different variables, we can generally accept that

$$\left| \underline{\underline{e}} \right| \sim \left| \underline{\underline{\omega}} \right| \sim \varepsilon, \; 0 < \varepsilon \ll 1 . \quad (2.8)$$

We will also use the kinematical and dynamic variables scaled with the parameter $\varepsilon$ as follows:

$$\underline{\underline{e}} \to \underline{\underline{e}} / \varepsilon, \quad \underline{\underline{\omega}} \to \underline{\underline{\omega}} / \varepsilon, \quad \underline{\underline{\sigma}} \to \underline{\underline{\sigma}} / \varepsilon, \quad f \to f / \varepsilon^2 . \quad (2.9)$$

In order to simplify analysis we introduce a special (generally local) Cartesian coordinate system $\{\hat{\underline{x}}\}$, whose one axis, say $\hat{x}_1$, is directed along the axis of initial director, $\underline{n}$. In $\{\hat{\underline{x}}\}$, where $\underline{n} = \{1,0,0\}$, the free energy (2.4) is represented as:

$$\hat{f} = (1/2G_0 + G_1 + G_2)\hat{e}_{11}^2 + 1/2G_0(\hat{e}_{22}^{a1} + e_{33}^2 + 2e_{23}^2)$$
$$+ \sum_{k=2,3}[(G_0 + G_1)\hat{e}_{1k}^{a1} + 2G_3\omega_{1k}\hat{e}_{1k}^{a1} + G_4\omega_{1k}^2]. \qquad (2.10)$$

Hereafter all the tensor components in $\{\hat{\underline{x}}\}$, scaled as shown in (2.9), are marked by upper caps.

The CE's (2.6a,b) written in $\{\hat{\underline{x}}\}$ and scaled as in (2.9) take the form:

$$\hat{\sigma}_{22}^{a1} = G_0 e_{22}; \quad \hat{\sigma}_{33}^{a1} = G_0 e_{33}; \quad \hat{\sigma}_{23}^{a1} = G_0 e_{23} \qquad (2.11a)$$

$$\hat{\sigma}_{11}^{a1} = (G_0 + 2G_1 + 2G_2)e_{11} \qquad (2.11b)$$

$$\begin{cases} \hat{\sigma}_{1k}^{a1} = (G_0 + G_1)e_{1k} + G_3\omega_{1k} \\ \hat{\sigma}_{1k}^{a1} = G_3 e_{1k} + G_4\omega_{1k} \end{cases} \quad (k = 2,3) \qquad (2.11c)$$

Here the CE's for anti-symmetric stress components are written only for $k < j$.

We call each of nine pairs $\{\hat{e}_{ij}^{a1}, \omega_{ij}\}$ of kinematic variables in (2.11) the $\{i,j\}$ deformational/rotational mode or simply $\{i,j\}$ *mode*. The modes in CE's (2.11a), which depend only on isotropic modulus $G_0$, are called *isotropic*, and all other in (2.11b,c) *nematic*. The nematic modes $\{1,1\}$ in CE (2.11b) and $\{1,k\}/\{k,1\}$ in (2.11c) are naturally called *stretching* and *shearing*, respectively. For simplicity we denote in the following the shearing nematic modes as $\{1,k\}$.

The thermodynamic stability constraints are found when demanding the quadratic form in (2.10) to be positive definite. Since all terms in (2.10), including two last quadratic forms for $k = 2$, $3$ $k = 3$, are independent, the (necessary and sufficient) stability conditions are:

$$G_0 > 0; \quad G_4 > 0; \quad G_0 + G_1 > 0; \quad 1/2G_0 + G_1 + G_2 > 0; \quad (G_0 + G_1)G_4 > G_3^2. \quad (2.12)$$

Combining (2.5) and (2.12) yields known stability conditions for coefficients in (2.3):

$$\mu_0 > 0; \quad \mu_1 > 0; \quad \mu_2 > 0; \quad 4\alpha_1\mu_2 > \alpha_2^2 \quad (\alpha_1 > 0). \qquad (2.13)$$

When the set of parameters $P = \{G_0,...,G_4\}$ satisfies (2.7) and (2.12), there are always 1-1 dependences between the tensors $\underline{e}$, $\underline{\omega}$ and $\underline{\sigma}$ described by CE's (2.6). This fact is evident for CE's (2.11a,b). It is also clear for CE's (2.11c) if they are considered as a coupled set of two linear algebraic equations with positive determinant,

$$D \equiv (G_0 + G_1)G_4 - G_3^2 > 0 . \qquad (2.14)$$

This result proved for particular case (2.11) in $\{\hat{x}\}$, is clearly valid for the general form of CE's (2.6).

## 3. Marginal stability, soft and semi-soft modes.

The marginal stability occurs when some inequalities in (2.12) are changed for equalities. Due to (2.7), only two marginal stability constraints can exist,

$$1/2 G_0 + G_1 + G_2 = 0, \qquad (3.1)$$

$$D \equiv (G_0 + G_1)G_4 - G_3^2 = 0 . \qquad (3.2)$$

We call the $\{i, j\}$ nematic modes whose parameters satisfy (3.1), (3.2) or both, *marginally stable.*

It is also possible to consider the case when stability conditions (2.12) are satisfied, but two last inequalities in (2.12) are close to those in (3.1) and (3.2):

$$1/2 G_0 + G_1 + G_2 = G_0 O(\delta), \qquad (3.1a)$$

$$D \equiv (G_0 + G_1)G_4 - G_3^2 = G_3^2 O(\delta) . \qquad (3.2a)$$

Here $0 < \delta << 1$. We call the $\{i, j\}$ stable nematic modes whose parameters $P$ satisfy (3.1a), (3.2a) or both, *nearly marginally stable.*

A particular $\{i, j\}$ mode is called *soft* if $\hat{\sigma}_{ij} = 0$ while $\partial \dot{e}_{ij}^{-1} \omega_{ij} = O_{ij}(1)$. CE's (2.11) show that the soft modes, if exist, are nematic, i.e. are described by CE's (2.11b,c). Due to (2.14) and (3.2) there are 1-1 dependences between relative rotations and respective strains in the shearing soft modes, although these strains are not unique. Evidently, these possible modes do not contribute in the free energy. It means that if both stretching $\{1,1\}$ and shearing $\{1, k\}$ modes exist, they deliver a minimum of the free energy in the parametric "space".

The *semi-soft* modes are defined as stable nematic modes $\{i, j\}$ in CE (2.11b,c), which for small positive numerical parameter $\delta$ satisfy the relations

$$\hat{\sigma}_{11} / G_0 = O(\delta) \text{ and } \hat{\sigma}_{1k} / G_3 = O(\delta), \text{ while } \partial \dot{e}_{ij}^{-1} \omega_{ij} = O_{ij}(1) . \qquad (3.3)$$

The stresses for semi-soft modes, as well as their contribution in free energy, are considerably smaller than for other modes. A nonlinear analysis of semi-soft deformations in terms of Warner potential [15] shows a good agreement with experimental data. This analysis is, however, outside the scope of our linear approach.

The general behavior of weakly elastic nematic solids can now be formally classified in terms of values of parameter $\delta$, as soft ($\delta = 0$), semi-soft ($0 < \delta << 1$), and hard ($\delta >\sim 1$).

We now establish relations between the existence of soft (or semi-soft) modes and equations (3.1), (3.2) (or (3.1a), (3.2a)). Constitutive equation (2.11b) shows that *the stretching nematic mode {1,1} is soft (semi-soft) if and only if it is marginally (or nearly marginally) stable.*

The shearing soft (semi-soft) nematic modes $\{1, k\}$ are trivially analysed using CE's (2.11c). If the shear stress components, $\hat{\sigma}_{1k}^{a\dashv} = \sigma_{1k}^{a} = 0$, the non-trivial solution of CE's (2.11c), the functions $\hat{e}_{1k}$ and $\hat{\omega}_{1k}$, exist only if the determinant $D = 0$, i.e. relation (3.2) is satisfied. It means that the shear soft modes $\{1, k\}$ are marginally stable. On the contrary, if $D = 0$, non-trivial solutions of (2.11c), $\hat{e}_{1k}$ and $\hat{\omega}_{1k}$, exist not only when $\hat{\sigma}_{1k}^{a\dashv} = \sigma_{1k}^{a} = 0$, but also when $\hat{\sigma}_{1k}^{a\dashv} \sim \sigma_{1k}^{a} \neq 0$. It means that the soft marginally stable $\{1, k\}$ modes are not unique, but along with the soft modes, non-soft but still marginally stable shear modes exist. As seen from (2.10), the soft shear modes are physically preferable as less energetically costly. Elementary analysis shows that the same situation holds for the semi-soft shearing modes too. This analysis is summarized as: *the soft (or semi-soft) $\{1, k\}$ shearing modes exist if and only if the marginal (or nearly marginal) stability condition (3.2) (or (3.2a)) is satisfied.*

## 4. Rotational (nearly rotational) invariance of stresses and soft (semi-soft) modes.

Until now, the Cartesian axes $\hat{x}_2^{\dashv}, x_3$ were assumed to be fixed in the plane $\{\hat{x}_2^{\dashv} x_3\}$ orthogonal to the director. We now consider the effect of rigid rotations of the axes $\hat{x}_2^{\dashv}, x_3$ in the plane $\{\hat{x}_2^{\dashv} x_3\}$ on stress behavior and related behavior of the soft and semi-soft modes. The orthogonal matrix $\underline{\underline{q}}$ which describe these plane rotations is:

$$q(\alpha) = \begin{pmatrix} 1 & 0 & 0 \\ 0 & \cos\alpha & -\sin\alpha \\ 0 & \sin\alpha & \cos\alpha \end{pmatrix}. \tag{4.1}$$

Denoting the stress tensor transformed with matrix $\underline{\underline{q}}$ as $\underline{\underline{\hat{\sigma}}}' = \underline{\underline{q}}^T \cdot \underline{\underline{\sigma}} \cdot \underline{\underline{q}}$, $\underline{\underline{\hat{\sigma}}}'^T = \underline{\underline{q}}^T \cdot \underline{\underline{\sigma}}^T \cdot \underline{\underline{q}}$ and calculating the transformed stress components $\hat{\sigma}'_{11}$, $\hat{\sigma}'_{1k}$ and $\hat{\sigma}'_{k1}$ ($k = 2,3$) yields:

$$\hat{\sigma}'_{11} = \sigma_{11}, \quad \hat{\sigma}'_{12} = \sigma_{12}\cos\alpha + \sigma_{13}\sin\alpha, \quad \hat{\sigma}'_{13} = -\sigma_{12}\sin\alpha + \sigma_{13}\cos\alpha. \tag{4.2}$$

The same relations hold for $\hat{\sigma}'_{k1}$. These formulae show that if for particular locations of coordinates $x'_2, x'_3$ in the plane $\{x'_2 x'_3\}$ the modes $\{1,k\}$ are soft, i.e. $\{\hat{\sigma}_{1k}\} = 0$ ($k = 2, 3$), they are soft for any Cartesian coordinates $x'_2, x'_3$ located in the same plane. We define the shear stress components as *rotationally invariant* when they do not change their values with the rotations described by (4.1). Then we can state that the shear stresses in soft modes are rotationally invariant. Relations (4.2) for the semi-soft modes where $\{\hat{\sigma}_{1k}\} = O(\delta)$ ($k = 2, 3$), show that the semi-soft modes are also rotationally invariant. Thus we obtained that *the set of soft (semi-soft) shearing modes is of continuum power.* This result known earlier from Olmsted paper [17] is almost evident for soft modes because the axes $x'_2, x'_3$ are located arbitrarily in the plane $\{x'_2 x'_3\}$. As seen from the first relation in (4.2), the soft stretching mode $\{1,1\}$ also remains to be soft (semi-soft) under the transformation with matrix (4.1).

We now consider the inverse problem: under which condition the stress components $\hat{\sigma}_{ij}$ are rotationally invariant? Once again, only plane rotations about the director (around axis $\hat{x}_1$) described by the matrix (4.1) are of interests here. In this case the formal definition of the rotationally invariant stress components $\hat{\sigma}_{ik}$ in (2.11) (if they exist) under orthogonal transformation (4.1) is:

$$\forall \alpha : \hat{\sigma}'_{ik} \equiv (\underline{\underline{q}}^T \cdot \underline{\underline{\sigma}} \cdot \underline{\underline{q}})_{ik} = \hat{\sigma}'_{ik}, \quad \sigma^{T'}_{ik} \equiv (\underline{\underline{q}}^T \cdot \underline{\underline{\hat{\sigma}}}'^T \cdot \underline{\underline{q}})_{ik} = \sigma^T_{ik}. \tag{4.3}$$

Using (4.2) we find that the only possible rotationally invariant stress components under constraint (2.7) are:

$$\hat{\sigma}'_{11} \equiv \sigma_{11}, \quad \hat{\sigma}'_{12} \equiv \sigma_{12}\cos\alpha + \hat{\sigma}'_{13}\sin\alpha = \sigma_{12}, \quad \hat{\sigma}'_{13} \equiv -\sigma_{12}\sin\alpha + \hat{\sigma}'_{13}\cos\alpha = \sigma_{13}. \tag{4.4}$$

And the same relations hold for $\hat{\sigma}_{k1}$. The solution of coupled second and third equations in (4.4), valid for any $\alpha$, is: $\hat{\sigma}_{12}^{-1} = \sigma_{21} = \hat{\sigma}_{13}^{-1} = \sigma_{31} = 0$. If the right-hand sides in (4.4) are changed for $\hat{\sigma}_{1k}(1 + O(\delta))$, $\hat{\sigma}_{1k}^{T}(1 + O(\delta))$, where $\delta$ is a small positive parameter, it is easily proved that this case describes "nearly" rotationally invariant shear stress modes responsible for occurring the semi-soft behaviour.

All the results proved above in this Section can be formulated as the *theorem*: *The shear stress components $\hat{\sigma}_{1k}$ and $\hat{\sigma}_{k1}$ (k = 2,3) in CE's (9c) are rotationally (or nearly rotationally) invariant if and only if the modes {1,k} are soft (semi-soft); the set of soft (semi-soft) shearing modes having the power of continuum.* It is clear that for free energy function (2.4), the rotational invariance is valid if and only if the marginal stability condition (3.2) (or nearly marginal stability condition (3.2a)) is satisfied. As seen from (4.2), (4.4), the rotationally (nearly rotationally) invariant stretching nematic mode {1,1} is not necessarily soft (semi-soft). It is soft (semi-soft) only if it is marginally (nearly marginally) stable. The formulated rotational invariance theorem is close to the rotational invariance principle postulated by Olmsted [17].

## 5. Renormalized equations for describing soft modes.

When the shearing modes {1,k} are soft the general solution of problems for CE's (2.6) is complicated by the fact that the shear strains $\hat{e}_{1k}$ in the soft modes are uncertain. Nevertheless, there is another important feature of the shearing soft modes, which makes the solution of problems for CE's (2.6) much easier than in general stable case. CE's (2.11c) show that when the shearing modes {1, k} are soft, the anti-symmetric components of stress tensor in $\{\hat{\underline{x}}\}$ disappear, because $\hat{\sigma}_{1k}^{a} = 0$. It means that in general $\underline{\underline{\sigma}}^{a} = 0$ for the weakly nematic solid with shearing soft modes, i.e. *if the shearing modes {1, k} are soft, the stress tensor is symmetric.*

Using the condition $\underline{\underline{\sigma}}^{a} = 0$ in (2.6b) yields:

$$\underline{\underline{\omega}} = (G_3 / G_4)(\underline{\underline{e}} \cdot \underline{n}\underline{n} - \underline{n}\underline{n} \cdot \underline{\underline{e}}).\qquad(5.1)$$

This relation has been found in paper [9], using different argument. Substituting (5.1) into (2.4) and (2.6a) results in the expressions for the renormalized free energy $f^r$ and symmetric extra stress $\underline{\underline{\sigma}}^r$:

$$f^r = 1/2 G_0 \left|\underline{\underline{e}}\right|^2 + G_1^r \, \underline{n}\underline{n} : \underline{\underline{e}}^2 + G_2^r (\underline{n}\underline{n} : \underline{\underline{e}})^2 \tag{5.2}$$

$$\underline{\underline{\sigma}}^r = G_0 \underline{\underline{e}} + G_1^r (\underline{n}\underline{n} \cdot \underline{\underline{e}} + \underline{\underline{e}} \cdot \underline{n}\underline{n}) + 2 G_2^r \underline{n}\underline{n}(\underline{\underline{e}} : \underline{n}\underline{n}) \quad (= \partial f^r / \partial \underline{\underline{e}} \, !) \, . \tag{5.3}$$

$$G_1^r = G_1 - G_3^2 / G_4; \quad G_2^r = G_2 + G_3^2 / G_4 \, . \tag{5.4}$$

Note that this renormalized formulation could still be roughly valid when the condition $\underline{\underline{\sigma}}^a = 0$ holds only approximately, as in the case of semi-soft shearing modes. In this case, the stability conditions for the free energy function (5.2) should be re-established anew. When the condition of marginal stability (3.1) holds, relations (23)-(25) are simplified to:

$$f^r / G_0 = 1/2 \left|\underline{\underline{e}}\right|^2 - \underline{n}\underline{n} : \underline{\underline{e}}^2 + 1/2(1 + \beta)(\underline{n}\underline{n} : \underline{\underline{e}})^2 \tag{5.2a}$$

$$\underline{\underline{\sigma}}^r / G_0 = \underline{\underline{e}} - (\underline{n}\underline{n} \cdot \underline{\underline{e}} + \underline{\underline{e}} \cdot \underline{n}\underline{n}) + (1 + \beta)\underline{n}\underline{n}(\underline{\underline{e}} : \underline{n}\underline{n}) \, . \tag{5.3a}$$

$$\beta = (G_0 + 2G_1 + 2G_2) / G_0 \, . \tag{5.4a}$$

Due to the third inequality in (2.12) parameter $\beta$ is positive if the stretching nematic mode {1,1} in (11b) is stable. When this stretching mode is also soft, $\beta = 0$, and the function $f^r$ in (5.2a) is minimized. Thus the presence of both stretching and shearing nematic modes brings the free energy $f$ to the minimum in the parametric space {P}. In this case the expressions (5.2a) and (5.3a) for renormalized free energy and stress, scaled with isotropic modulus $G_0$, have no parameters at all, which simplifies the general analysis of deformations.

We now demonstrate simple invariant expressions of strains for shearing and stretching soft modes, close to that considered in paper [9]. These soft modes can be easily found from the CE (5.3a).

The strain and relative rotation for shearing soft mode that satisfies CE (5.3a) are:

$$\forall \underline{c}, \; \underline{c} \cdot \underline{n} = 0: \; \underline{\underline{e}}^s = \gamma(\underline{c}\underline{n} + \underline{n}\underline{c}), \quad \underline{\underline{\omega}}^s = \gamma(G_3 / G_4)(\underline{c}\underline{n} - \underline{n}\underline{c}) \, . \tag{5.5}$$

Here $\gamma$ is the factor characterizing intensity of shear, and $\underline{c}$ is an arbitrary unit vector orthogonal to the director. Relations (5.6) demonstrate the fact of existing the continuous set of soft shear modes. It is easy to see that substituting (5.5) into (5.3a) nullifies the stress, i.e. the mode described by (5.5) is indeed soft.

The strain tensor describing stretching soft mode is presented as:

$$\underline{\underline{e}}^e = \varepsilon(\underline{nn} - \frac{1}{3}\underline{\underline{\delta}}). \qquad (5.6)$$

Here $\varepsilon$ is the factor characterizing intensity of stretching in diction of director, and $\underline{\underline{\delta}}$ is the unit tensor. When $\beta = 0$, the stretching tensor (5.6) nullify the components $\sigma_{11}^r$ and $\sigma_{22}^r = \sigma_{33}^r$ in CE (5.3a). It means that the nematic elastic bar extended along the director with stretching soft mode will need no stresses for this type of deformation. Although $\underline{\underline{\sigma}}^r$ in (5.3a) is the extra stress tensor, the pressure $p$ in this case is equal to zero, because it satisfies the free-stress boundary condition. Evidently, the soft mode strains described by (5.5) and (5.7) are volume preserving since $e_{ii} = 0$ in both cases. It should also be noted that due to the linear character of CE (5.3a), any linear combination of $\underline{\underline{e}}^s$ and $\underline{\underline{e}}^e$ presents a combined shear-stretch soft mode when $\beta = 0$.

## 6. Example: weak Warner potential.

The general expression for this potential was derived by Warner and co-workers [11,15] for nematic elastomers, using entropy concept Assuming that the parameters of anisotropy $l_\parallel$ and $l_\perp$ are not changed after applying strains, Olmsted [17] obtained the expression for the weak Warner potential, which in our notations has the form:

$$\frac{f^w}{G_o} = 1/2\left|\underline{\underline{e}}^2\right| + \frac{(\Delta l)^2}{4l_\perp l_\parallel}[\underline{nn}:\underline{\underline{e}}^2 - (\underline{nn}\cdot\underline{e})^2 - \underline{nn}:\underline{\underline{\omega}}^2] - \frac{\Delta l^2}{2l_\perp l_\parallel}\underline{nn}:(\underline{e}\cdot\underline{\omega}) ;$$
$$(\Delta l)^2 = (l_\parallel - l_\perp)^2; \quad \Delta l^2 = l_\parallel^2 - l_\perp^2. \qquad (6.1)$$

It is seen that (6.1) presents a particular version of the free energy function (2.4), where

$$G_1 = -G_2 = G_4 = G_0\frac{(l_\parallel - l_\perp)^2}{l_\parallel l_\perp}; \quad G_3 = G_0\frac{l_\parallel^2 - l_\perp^2}{l_\parallel l_\perp}. \qquad (6.2)$$

Then CE's (2.6) are valid for extra stress in the case of potential (6.1) with specification (6.2). It is also seen that the parameters (6.2) satisfy the marginal stability condition (3.2) for shearing modes $\{1, k\}$ along with other thermodynamic stability constraints in (12), i.e. *the shearing modes $\{1, k\}$ as described by the weak Warner potential are always soft.* Then the renormalized formulation (5.2a), (5.3a) always holds for weak Warner potential with the particular value $\beta = 1$. Because of this fact, the parameters (6.2) cannot describe the marginal stability constraint (3.1) for stretching mode {1,1}, i.e. this mode is not soft. A possible reason for that is that the initial parameters of anisotropy $l_{\parallel}$ and $l_{\perp}$ in (6.1) have been assumed unchanged after applying strains.

## 7. Conclusions.

This paper was motivated by the attempt of the present authors to apply the general nematic formulation for describing various polymer systems (not only soft nematic gels) in equilibrium and non-equilibrium situations (e.g. see [23,24]). Along with very soft nematic elastomers and gels there are also semi-soft and even hard nematic elastomers. Additionally, there are composites of liquid-like polymers, filled with attractive anisotropic (micro/nano scale size) particles, which above a percolation threshold in filler concentration demonstrate nematic elastic properties for relatively small stresses. The behaviour of these nematic systems with various degrees of softness cannot be simply explained in terms of broken isotropic symmetry [18].

The present paper analysed the existence and description of soft, semi-soft and harder modes for weakly elastic nematic solids employing a continuum approach proposed by de Genes [1], without using any additional physical assumptions. The analysis was trivialized when we used a special Cartesian coordinates $\{\hat{x}\}$ whose one axis is directed along the director. The following results were obtained:

1. The free energy (4) predicts the soft and semi-soft modes to exist if and only if the conditions of marginal (or nearly marginal) stability are satisfied. Along with shearing soft modes, potential (4) also predicts the existence of marginally stable extensional soft and semi-soft modes.

2. When marginally stable soft nematic shear modes exist, the stress tensor is symmetric. This fact results in a simplified description of soft shearing modes with a renormalized expression $f^r$ for the free energy. Scaled by the isotropic elastic (or high-elastic) modulus, the function $f^r$ contains only one, non-dimensional material parameter. It is shown that this renormalized formulation is always applicable to the weak version of the Warner potential, however, without description of possible stretching soft mode. The presence of both the extensional and shearing soft modes minimizes the reduced free energy so that there are no parameters in its formulation scaled by the isotropic modulus.

3. The theorem of rotational invariance proved in the paper establishes the equivalency between existence of shearing soft modes and shear stress rotational invariance. This theorem is close to the principle of rotational invariance postulated by Olmsted [16].

We finally should mention that for possible application of DG potential (2.4) to various materials (not soft gels), the moduli $G_k$ in (2.4) could depend differently on temperature. Therefore the marginal stability (or nearly marginal stability) constraints (3.1), (3.2) could generally exist only in certain temperature intervals.